\documentclass[aps,prd,nofootinbib,twocolumn,superscriptaddress,preprintnumbers]{revtex4}

\usepackage{amssymb}
\usepackage{amsmath}
\usepackage{epsfig}
\usepackage{hyperref}

\usepackage{amsbsy}

\makeatletter
\def\simgt{\mathrel{\lower2.5pt\vbox{\lineskip=0pt\baselineskip=0pt
           \hbox{$>$}\hbox{$\sim$}}}}
\def\simlt{\mathrel{\lower2.5pt\vbox{\lineskip=0pt\baselineskip=0pt
           \hbox{$<$}\hbox{$\sim$}}}}
\makeatother

\newcommand{\be}{\begin{equation}}
\newcommand{\ee}{\end{equation}}
\newcommand{\bea}{\begin{eqnarray}}
\newcommand{\eea}{\end{eqnarray}}
\newcommand{\Eq}[1]{Eq.~(\ref{#1})}

\newcommand{\Sec}[1]{Sec.~\ref{#1}}
\newcommand{\App}[1]{App.~\ref{#1}}
\newcommand{\Fig}[1]{Fig.~\ref{#1}}
\newcommand{\Ref}[1]{Ref.~\cite{#1}}
\newcommand{\Refs}[1]{Refs.~\cite{#1}}

\newcommand{\MPl}{M_{\rm Pl}}
\newcommand{\bPhi}{{\boldsymbol \Phi}}

\newcommand{\bs}[1]{\boldsymbol #1}

\newcommand{\vev}[1]{\left\langle #1 \right\rangle}
\newcommand{\nn}{\nonumber}

\newcommand{\Arg}{\mathop{\text{Arg}}}

\newcommand{\bK}{{\boldsymbol K}}
\newcommand{\bX}{{\boldsymbol X}}
\newcommand{\bT}{{\boldsymbol T}}
\newcommand{\bQ}{{\boldsymbol Q}}
\newcommand{\boldf}{{\boldsymbol f}}
\newcommand{\bOmega}{{\boldsymbol \Omega}}
\newcommand{\bW}{{\boldsymbol W}}

\begin{document}

\preprint{MIT-CTP 4244}

\title{The Spectrum of Goldstini and Modulini}

\author{Clifford Cheung}
\affiliation{Berkeley Center for Theoretical Physics, 
  University of California, Berkeley, CA 94720, USA}
\affiliation{Theoretical Physics Group, 
  Lawrence Berkeley National Laboratory, Berkeley, CA 94720, USA}

\author{Francesco D'Eramo}
\affiliation{Center for Theoretical Physics, 
  Massachusetts Institute of Technology, Cambridge, MA 02139, USA}
  
\author{Jesse Thaler}
\affiliation{Center for Theoretical Physics, 
  Massachusetts Institute of Technology, Cambridge, MA 02139, USA}

\begin{abstract}
When supersymmetry is broken in multiple sectors via independent dynamics, the theory furnishes a corresponding multiplicity of ``goldstini'' degrees of freedom which may play  a substantial role in collider phenomenology and cosmology.  In this paper, we explore the tree-level mass spectrum of goldstini arising from a general admixture of $F$-term, $D$-term, and almost no-scale supersymmetry breaking, employing  non-linear superfields and a novel gauge fixing for supergravity discussed in a companion paper.  In theories of $F$-term and $D$-term breaking, goldstini acquire a mass which is precisely  twice the gravitino mass, while the inclusion of no-scale breaking renders one of these modes, the modulino, massless.  We argue that the vanishing modulino mass can be explained in terms of an accidental and spontaneously broken ``global'' supersymmetry.
\end{abstract}

\maketitle

\section{Introduction}

If supersymmetry (SUSY) is indeed a symmetry of nature then it must be spontaneously broken.  While low energy phenomenology is largely independent of the particulars of SUSY breaking in the ultraviolet, a notable exception occurs if these dynamics yield additional light degrees of freedom.  For instance the gravitino, whose mass can range from the weak scale down to an electronvolt, is an intriguing dark matter candidate which is absolutely crucial for SUSY cosmology and collider phenomenology \cite{Wess:1992cp,Cremmer:1978hn,Martin:1997ns}.  In other instances, the spectrum may contain light $R$-axions \cite{Bagger:1994hh, Goh:2008xz} or pseudomoduli \cite{Shih:2009he} which arise from SUSY breaking and  impact low energy physics.

Recently, it has been shown that ``goldstini'' may also appear in the low energy spectrum \cite{Cheung:2010mc, Cheung:2010qf}.  These states arise if there exist a multiplicity of sectors which each break SUSY via independent dynamics.  For $N$ such sectors, the spectrum is comprised of a gravitino with mass $m_{3/2}$ (whose longitudinal mode is a ``diagonal'' goldstino eaten via the super-Higgs mechanism) and $N-1$ uneaten goldstini.  In the simplest scenario, these uneaten goldstini acquire a universal tree-level mass of $2m_{3/2}$ \cite{Cheung:2010mc} due to supergravity (SUGRA) effects.  Interesting variations  arise in the context of strong dynamics or equivalently extra-dimensional warping \cite{Craig:2010yf}, as well as in cases with less sequestering \cite{Argurio:2011hs}.   Specific implications for colliders \cite{Cheung:2010mc, Cheung:2010qf,Thaler:2011me}, dark matter \cite{Cheung:2010mc,Cheung:2010qf,Cheng:2010mw}, and model building \cite{McCullough:2010wf,Izawa:2011hi} have also been studied.  For an earlier incarnation of this idea in the context of brane worlds, see \Ref{Benakli:2007zza}.

The spectrum of goldstini is further enriched if SUSY breaking is directly tied to gravity, as in so-called ``almost no-scale'' models \cite{Luty:2002hj}.  In this case, a bizarre cancellation \cite{Cheung:2010mc} renders one linear combination of fermions---the modulino---massless at tree-level even after including SUGRA effects.\footnote{To our knowledge, this tree result was first derived in the appendix of \Ref{Cheung:2010mc}, though similar observations were made in \Ref{Rattazzi:1999qg}.  At one loop, one expects the modulino to acquire a mass of order $m_{3/2}/16\pi^2$.  }  
Because goldstini can have a dramatic impact on SUSY phenomenology, it is of utmost importance to understand their spectrum and interactions in the most general case of multiple SUSY breaking.

In this paper, we derive the spectrum of goldstini and modulini for a general theory of $F$-term, $D$-term, and almost no-scale SUSY breaking with the aid of a novel gauge fixing of SUGRA introduced in a companion paper \cite{goldstini3}.  In this gauge, it is possible to compute the spectrum and couplings of matter fields in the language of superspace without making any reference to the component form SUGRA Lagrangian.  Mechanically, the entire effect of this gauge fixing is to introduce a non-standard but easily manipulated conformal compensator which effectively decouples the graviton and gravitino from calculations relevant to the matter and gauge fields alone.  

The remainder of the paper is organized as follows.  In \Sec{sec:improved}, we review the improved compensator formalism of \Ref{goldstini3}.  We then compute the spectrum of goldstini for arbitrary $F$-term and $D$-term SUSY breaking sectors in \Sec{sec:goldstini}, with details left to the appendices.  In the minimal scenario, the goldstini acquire a universal tree-level mass equal to $2m_{3/2}$, and we study possible deviations arising from small linear terms in the K\"ahler potential.  We then extend our analysis to more general goldstini/modulini scenarios in \Sec{sec:modulini}, and find a concise and more physical explanation for the tree-level massless fermion that arises in almost no-scale SUSY breaking models.  Namely, the modulino is revealed as a ``goldstino'' arising from the breaking of an accidental hidden ``global'' SUSY.  We conclude in \Sec{sec:conclusions}.

\section{The Improved Compensator}
\label{sec:improved}
 In the standard gauge fixing of conformal SUGRA \cite{conformalcompensator}, the conformal compensator takes the form $\bPhi = 1 + \theta^2 F_\Phi$,\footnote{Throughout, we  use the convention of \Ref{goldstini3} in which boldface ($\bX$) refers to superfields and regular typeface refers to component component fields ($X$).  Subscripts on functions refer to the corresponding field derivatives, i.e.\ $\Omega_i \equiv \partial \Omega / \partial X^i$ and $\Omega_{\bar{i}} \equiv \partial \Omega / \partial X^{\dagger \bar{i}}$, and indices are raised and lowered using the K\"{a}hler metric.} and the SUGRA action is
\begin{align}
\label{eq:naiveconformalcompensator}
\mathcal{L}_{\rm SUGRA} &= -3 \int d^4 \theta \; \bPhi^\dagger \bPhi \; e^{-{\bs K}/3} + \int d^2 \theta \; \bPhi^3 \; {\bs W} + \mathrm{h.c.} \nn \\
& \qquad ~ + \frac{1}{4} \int d^2 \theta \;  {\bs f}_{ab} {\bs W}^{a\alpha} {\bs W}^{b}_{\alpha} + \mathrm{h.c.} + \ldots,
\end{align}
where the ellipsis ($\ldots$) indicates additional  terms involving the graviton, gravitino, and vector auxiliary field, and  we  work in natural units where $\MPl = 1$.  Note that the terms in the ellipsis do not take a simple form in terms of superfields, and are naturally expressed in terms of component fields.
 Here the K\"{a}hler potential ${\bK}$ is a function of the chiral and vector superfields, and the superpotential ${\bs W}$ and gauge kinetic term ${\bs f}_{ab}$ are holomorphic functions of the chiral superfields alone.  The chiral superfields are defined in components as
\be
\label{eq:Xpara}
{\bs X}^i = X^i + \sqrt{2} \theta \chi^i + \theta^2 F^i,
\ee
where we work in a basis of vanishing vacuum expectation value (vev) for each field, $\langle X^i\rangle =0$.
As discussed in our companion paper \cite{goldstini3}, the naive application of \Eq{eq:naiveconformalcompensator} with $\bPhi = 1 + \theta^2 F_\Phi$ yields incorrect answers unless the  terms denoted by the ellipsis are carefully included.  These additional terms properly account for essential mixing terms between the gravity multiplet and matter fields.

The key result of \Ref{goldstini3} is that there exists an improved gauge fixing for the conformal compensator which, for all intents and purposes, decouples  the gravity multiplet from calculations relevant to the matter fields alone.  In particular, the choice
\bea
\label{eq:phicorrect}
\bPhi &=&    
e^{{\bs Z}/3}
(1+ \theta^2 F_{ \Phi} ), \\
{\bs Z} &=&  \left\langle K/2 - i \Arg W \right \rangle + \langle K_i \rangle {\bs X}^i,
\eea
removes the undesirable graviton and gravitino mixing terms.  With this variant of the conformal compensator, one can use \Eq{eq:naiveconformalcompensator} and justifiably ignore the terms in the ellipsis.  Moreover,  \Eq{eq:naiveconformalcompensator} is expressed entirely in terms of superfields---without reference to component fields---so calculations are conveniently amenable to various superspace tricks.  This will be extremely useful later on, when we compute the mass spectrum of goldstini.  This gauge fixing has the added bonus that 
\be
\label{eq:Fphivev}
\langle F_\Phi \rangle = m_{3/2}
\ee
after adjusting the cosmological constant to zero, so it is straightforward to identify the dependence of physical quantities on the gravitino mass. Note that this gauge fixing can be alternatively understood as a prescient K\"aher transformation
\bea
{\bs K} &\rightarrow& {\bs K} - {\bs Z}- {\bs Z}^\dagger, \\ 
{\bs W} &\rightarrow & e^{{\bs Z}}{\bs W },
\eea
which removes unwanted linear terms in the K\"ahler potential.  Hence, this gauge fixing effectively converts the no-scale component of SUSY breaking into $F$-term SUSY breaking.

For theories comprised of multiple sectors which are sequestered from each other, it is  convenient to describe the physics in terms of  $\bOmega \equiv -3 \exp({-\bK/3})$, where $\bOmega$ is a sum of contributions from each sector.  Using the fact that $\vev{K_i} = -3 \vev{\Omega_i/\Omega}$, the preferred gauge fixing takes the form
\be
\bPhi = \sigma_0 \exp \left[ -\frac{\vev{\Omega_i} {\bs X}^i}{\vev{\Omega}}  \right] (1+ \theta^2 F_{ \Phi} ),
\ee
where
\be
\label{eq:sigma0vev}
\sigma_0 = \vev{\frac{W^*}{W}}^{1/6} \vev{\frac{-3}{\Omega}}^{1/2}.
\ee
It is also sometimes helpful to expand $\bPhi$ in components,
\be
\label{eq:phicomponents}
\bPhi = \sigma_0 \exp \left[ -\frac{\vev{\Omega_i} {X}^i}{\vev{\Omega}}  \right]   \left(1 - \sqrt{2} \theta  \frac{\vev{\Omega_i} \chi^i}{\vev{\Omega}} + \theta^2 \widetilde{F}_\Phi \right),
\ee
where
\be
\label{Ftildedef}
\widetilde{F}_\Phi \equiv \bPhi|_{\theta^2} = F_\Phi - \frac{\vev{\Omega_i}}{\vev{\Omega}} F^i.
\ee
Here $F_\Phi$ can be thought of as the contribution to SUSY breaking from $F$-terms and $D$-terms.
Finally note that by applying a constant K\"ahler transformation, one can always adjust $\Arg \vev{W} = 0$ and $\vev{\Omega} = -3$, resulting in $\sigma_0 = 1$ and simplifying the expression for $\bPhi$.  

\section{$\bs F$-term and $\bs D$-term Goldstini}
\label{sec:goldstini}

Let us now consider the spectrum of goldstini in a general theory of $F$-term and $D$-term SUSY breaking.  As we will see, goldstini in $F$-term theories have a mass of $2m_{3/2}$, and this curious factor of 2 persists  in the presence of $D$-term SUSY breaking.  Afterwards, we will employ the variant conformal compensator in \Eq{eq:phicorrect}  to understand small deformations away from the pure $F$-term and $D$-term breaking limit.

\subsection{Review of Goldstini}

The premise of the goldstini framework is that SUSY is broken independently in $N$ sequestered sectors \cite{Cheung:2010mc}.  A priori, the superfields of the visible sector can be coupled via non-gravitational interactions to zero, one, or more than one of these SUSY breaking sectors.  As discussed at length in \Ref{Cheung:2010mc}, if the visible sector couples non-gravitationally to zero sectors (as in anomaly mediation) or just to one sector, then the visible sector has little effect on the goldstini dynamics, since there still effectively exist $N$ sequestered sources of SUSY breaking.  However, if the visible sector couples to more than one SUSY breaking sector, then it can mediate large effects between the SUSY breaking sectors, inducing possibly significant modifications to the goldstini spectrum and couplings \cite{Cheung:2010mc,Argurio:2011hs}.  In what follows, we will assume that this is not the case and that the visible sector couples to no more than one SUSY breaking sector.

 In the global SUSY limit, each sector contains a corresponding massless goldstino.  Including SUGRA effects, one linear combination of the goldstini is eaten via the super-Higgs mechanism, leaving the remaining $N-1$ uneaten goldstini  in the spectrum as physical degrees of freedom.  An analogous effect occurs in the standard model, since  the Higgs sector and the QCD sector independently break electroweak symmetry, giving rise to two sets of Nambu-Goldstone bosons.  One linear combination is eaten to form the longitudinal components of the $W^\pm$/$Z^0$ bosons, while the orthogonal combination are the standard model pions $\pi^\pm/\pi^0$.

In the minimal goldstini scenario, each sector is completely sequestered from the other,  so the SUGRA potentials take a special form
\bea
\label{eq:sequesteredSUGRA}
\bOmega &=& -3 + \sum_A \bOmega^{A},\\
\bW &=& m_{3/2}  + \sum_A \bW^{A},\nonumber \\
\boldf_{ab} &=& \sum_A \boldf_{ab}^A, \nonumber
\eea
where $A=1,2,\ldots,N$ is an index labeling the various sectors.  In more general cases, the $N$ sectors are not perfectly sequestered, but as shown in \Ref{Cheung:2010mc,Argurio:2011hs}, it is typical for such mixings to be loop-suppressed.\footnote{Roughly speaking, the absence of sequestering induces corrections to the goldstini masses of the order $\delta m_\eta \simeq \widetilde{m}_{\rm soft} / (16 \pi^2)^n$, where $\widetilde{m}_{\rm soft}$ are visible sector soft masses and $n$ is the number of visible sector loops needed to connect two different hidden sectors \cite{Cheung:2010mc}.  In the context of multiple gauge mediation, this correction was calculated explicitly in \Ref{Argurio:2011hs}.}  As alluded to in the text below \Eq{Ftildedef}, in \Eq{eq:sequesteredSUGRA} we have assumed that $\vev{\Omega^A} = 0$, $\vev{W^A} = 0$, and $m_{3/2}$ is real, implying that $\vev{\Omega} = -3$ and $\Arg\vev{W} = 0$.  This results in a simplified form for the conformal compensator
\be
\bPhi =   e^{\vev{\Omega_i}\bX^i/3} (1+ \theta^2 m_{3/2} ),
\ee
where we have used \Eq{eq:Fphivev} and assumed that the cosmological constant has been adjusted to zero.\footnote{Because we are only interested in calculating fermion masses, the replacement $F_\Phi \rightarrow m_{3/2}$ is justified, though in general $F_\Phi$ has additional scalar field dependence.}

The other assumption of the minimal goldstini scenario is that SUSY breaking is independent of gravitational dynamics.  Recall that holomorphic K\"ahler terms are unphysical in the limit of global SUSY, but can be very important when SUGRA effects are properly accounted for.  In particular, the vacuum structure can fundamentally change at finite $\MPl$, as in the case of no-scale SUSY breaking.  For the moment, let us sidestep this important subtlety  and assume that $\langle \Omega_i \rangle = 0$, which is to say that the K\"ahler potential does not contain linear terms in $\bX^i$ at the vacuum.  In such a scenario  SUSY breaking is intrinsically global, in the sense that it is preserved when $\MPl\rightarrow\infty$.\footnote{Strictly speaking, we only need $\vev{\Omega_i}$ to be small compared to $\vev{\Omega_{i\bar{j}}}$.  See the appendix of \Ref{Cheung:2010mc}.}

Demanding that $\langle \Omega_i \rangle = 0$  simplifies our calculation because we can completely ignore the $e^{\vev{\Omega_i}\bX^i / 3} = 1$ term in $\bPhi$.  Thus, in order to compute the fermion mass spectrum in the minimal goldstini scenario, we can employ the standard conformal compensator often quoted in the literature, $\bPhi = 1 + \theta^2 m_{3/2}$.  This simplification will \emph{not} be valid for the modulini calculation in \Sec{sec:modulini}, though it will turn out to be approximately correct for the goldstini deformations in \Sec{sec:goldstinideformations}.

\subsection{Goldstini Masses}
\label{subsec:minimialmasscalc}

The spectrum of goldstini has already been calculated in \Ref{Cheung:2010mc} using $\bPhi = 1 + \theta^2 m_{3/2}$, but only for a simple theory in which each sector exhibits single field, Polonyi-type SUSY breaking.  Here, we  derive the goldstini spectrum for  arbitrary SUSY breaking sectors with $F$-terms and $D$-terms, and show that the tree-level goldstini masses are $2m_{3/2}$ provided that $\langle \Omega_i \rangle = 0$.  Because the conformal compensator method (in our preferred gauge fixing) is valid directly in superspace, we can compute the spectrum of goldstini purely in the language of superfields.  In particular, we will find a non-linear parametrization of superfields \cite{Komargodski:2009rz,Cheung:2010mc} to be particularly illuminating.

To begin, consider a single sector labeled by $A$.  While multiple fields in this sector can acquire non-zero $F$-terms and $D$-terms, we can always parameterize these fields, ${\bs X}^{Ai}$ and ${\bs V}^{Aa}$, in terms of the goldstino direction $\eta^A$ in each sector,
\begin{align}
\label{eq:nonlineardefs}
{\bs X}^{Ai} &= \left(\theta + \frac{1}{\sqrt{2}}\frac{\eta^A}{F^A_{\rm eff}}\right)^2 F^{Ai},  \\
{\bs V}^{Aa} & =  \left|\theta + \frac{1}{\sqrt{2}} \frac{\eta^A}{F^A_{\rm eff}}\right|^4 D^{Aa}, \nonumber
\end{align}
where for simplicity we have elided terms involving derivatives on the goldstino as well as the other physical degrees of freedom.  While this  parameterization may be somewhat unfamiliar, it can be easily understood as arising from a field redefinition, as shown in \App{app:details}.  

The goldstino decay constant for each sector is a positive number $F^A_{\rm eff}$ defined by
\be
\label{eq:goldstinodecayconstant}
(F^A_{\rm eff})^2 = F^{Ai} F^{*A\bar{j}} g_{i\bar{j}}  + \frac{D^{Aa} D^{Ab} f^A_{ab}}{2},
\ee
where $g_{i\bar{j}} = \langle \Omega_{i\bar{j}} \rangle$.\footnote{The identification of the K\"ahler metric with $\langle \Omega_{i\bar{j}} \rangle$ is only true because we are assuming $\langle \Omega_i \rangle = 0$, and have adjusted $\vev{\Omega} = -3$.}  After expanding out ${\bs X}^{Ai}$ and ${\bs V}^{Aa}$ and isolating their fermionic components, it is clear that $\eta^A$ does indeed correspond to the goldstino direction for sector $A$, since
\be
\eta^A = \frac{1}{F^A_{\rm eff}} \left( g_{i\bar{j}} F^{*A\bar{j}} \chi^{Ai} - \frac{i}{\sqrt{2}} f^A_{ab} D^{Ab} \lambda^{Aa} \right).
\ee
With the proper definition of the goldstino decay constant $F^A_{\rm eff}$, the goldstino $\eta^A$ is canonically normalized. 

We have chosen the novel parameterization in \Eq{eq:nonlineardefs} because this allows us to treat each sector as if it possesses an independent $\theta$ coordinate
\be
\label{eq:coordshift}
\theta^{A} = \theta + \frac{1}{\sqrt{2}} \frac{\eta^{A}}{F_{\rm eff}^{A}}.
\ee
With this coordinate shift, ${\bs X}^{Ai}$ and ${\bs V}^{Aa}$ are now \emph{independent} of the goldstino, and the only remaining goldstino dependence is in the conformal compensator, which can be suggestively rewritten as 
\be
\label{eq:PhiAshift}
\bPhi \equiv 1 + \left(\theta^{A} - \frac{1}{\sqrt{2}} \frac{\eta^{A}}{F_{\rm eff}^{A}}\right)^2 m_{3/2},
\ee
for any particular sector $A$.  For example, the Lagrangian for the matter fields in sector $A$ is given by
\begin{align}
\mathcal{L}^{A} &= \int d^4 \theta^{A} \, \bPhi^\dagger \bPhi \, {\bs \Omega}^{A} + \int d^2 \theta^A \, \bPhi^3 \, {\bs W}^{A} + \mathrm{h.c.}  \\
& \qquad ~ + \frac{1}{4} \int d^2 \theta^A {\bs f}^A_{ab}{\bs W}^{Aa\alpha} {\bs W}^{Ab}_{\alpha} + \mathrm{h.c.} + \ldots. \nonumber
\end{align}

Expanding $\mathcal{L}^{A}$ to quadratic order in the goldstino field, and extracting the $\eta^A$ mass term, we find
\be
\label{eq:almostthere}
\mathcal{L}^{A} \supset \frac{1}{2} m_{3/2} \left( \frac{ m_{3/2} {\bOmega}^{A}|_{\bar{\theta}^2} + {\bs \Omega}^{A}|_{\theta^4} + 3 \bW^{A}|_{\theta^2}}{(F_{\rm eff}^{A})^2} \right)  \eta^{A} \eta^{A}.
\ee
Note that
\be
\vev{{\bOmega}^{A}|_{\theta^2}} = \vev{\Omega^A_i F^{Ai}},
\ee
which is zero by assumption.  After solving the $F$-term and $D$-term equations of motion in sector $A$ (see \App{app:details})
\be
\label{eq:altFrelationship}
{\bOmega}^{A}|_{\theta^4} + 3 \bW^{A}|_{\theta^2} = -2 (F_{\rm eff}^{A})^2,
\ee
yielding
\be
-\frac{1}{2} (2m_{3/2}) \eta^{A} \eta^{A}.
\ee

Thus, we have demonstrated that each goldstino has a tree-level mass of $2m_{3/2}$.  One linear combination of the goldstini corresponds to the true goldstino which is eaten to become the longitudinal mode of the gravitino,
\be
\eta_{\rm eaten} = \sum_A \frac{F_{\rm eff}^A}{F_{\rm eff}} \eta^A, \qquad
(F_{\rm eff})^2 = \sum_A (F_{\rm eff}^A)^2.
\ee
Since the goldstino mass matrix is diagonal, isolating the eaten goldstino does not affect the mass spectrum of the uneaten goldstini.  Therefore, we recover our result that the uneaten goldstini have a universal tree-level mass given by
\be
\label{eq:finalgoldstinomass}
m_\eta = 2m_{3/2}.
\ee
This formula holds for arbitrary $F$-term and $D$-term SUSY-breaking sectors, provided that $\langle \Omega_i \rangle = 0$.

\subsection{Goldstini Deformations}
\label{sec:goldstinideformations}

The scenario of $F$-term and $D$-term SUSY breaking is a convenient starting point from which to understand the dynamics of multiple SUSY breaking.  However, there are many possible deformations away from this canonical setup, of which a number have been explored in \Refs{Craig:2010yf,Argurio:2011hs}.  Given our discussion in \Sec{subsec:minimialmasscalc}, the most obvious departure from this very simplest theory occurs when $\langle \Omega_i \rangle \not= 0$.  In what follows, we explore the physics corresponding to small perturbations away from $\langle \Omega_i \rangle = 0$, leaving a discussion of large values of $\vev{\Omega_i}$ for \Sec{sec:modulini}.

The calculation in  \Sec{subsec:minimialmasscalc} is modified in two important ways when $\langle \Omega_i \rangle \not= 0$.  It is easiest to understand these two differences  by expanding the conformal compensator in components as in \Eq{eq:phicomponents},
\be
\bPhi = e^{\vev{\Omega_i} X^i/3} \left(1 + \sqrt{2} \theta  \frac{\vev{\Omega_i} \chi^i}{3} + \theta^2 \widetilde{F}_\Phi \right)
\ee
where we have again used a K\"ahler transformation to adjust $\Arg \vev{W} = 0$ and $\vev{\Omega} = -3$.  Here, the highest component of $\bPhi$ is
\be
\widetilde{F}_\Phi \equiv \bPhi|_{\theta^2} = m_{3/2} + \frac{\vev{\Omega_i}}{3} F^i,
\ee
after adjusting the cosmological constant to zero.

The first difference is that the highest component of $\bPhi$ is no longer equal to $m_{3/2}$, but rather  $\widetilde{F}_\phi$.  Thus, when applying the manipulation in \Eq{eq:PhiAshift} for this theory, we find that the uneaten goldstino mass is proportional to $\widetilde{F}_\Phi$ rather than $m_{3/2}$.\footnote{There are subtleties in this statement that we will encounter in \Sec{sec:modulini} relating to the diagonalization of the goldstino mass matrix.}  This effect was studied in some detail in \Ref{Craig:2010yf}, and simply rescales the  goldstini mass spectrum, leaving the gravitino mass fixed. 

The second difference is that there is an additional $\vev{\Omega_i}\chi^i$ term in the fermionic component of $\bPhi$.  This  complication is not visible in the naive parametrization of $\bPhi$, and only appears when using the improved gauge fixing proposed in \Ref{goldstini3}.  To understand the effect of this term in a concrete setting, consider two SUSY breaking sectors labelled by $A=1,2$ with $F_{\rm eff}^1 \gg  F_{\rm eff}^2$.  To ensure small perturbations away from the pure $F$-term and $D$-term SUSY breaking limit, we will assume that sector 1 is of the form discussed in \Sec{subsec:minimialmasscalc}, such that $\vev{\Omega_i^1} = 0$ and $\widetilde{F}_\Phi \simeq m_{3/2}$.  We then assume for simplicity that sector 2 is comprised of a single chiral field $\bX$ with $R$-charge 2 with a K\"ahler and superpotential 
\be
\label{eq:BXLagrangian}
\bOmega = \bX^{\dagger} \bX - \frac{(\bX^\dagger \bX)^2}{M^2}, \qquad \bW = f \bX.
\ee
Here, we are using a standard linear parametrization for $\bX$ and will allow $\vev{X} \not=0$.  Since $F_{\rm eff}^1 \gg  F_{\rm eff}^2$, the uneaten goldstino $\eta$ can be identified with the fermionic component of $\bX$ up to $F_{\rm eff}^2/F_{\rm eff}^1$ corrections which can be justifiably ignored.

In the absence of SUGRA effects, $\langle X\rangle  =\vev{\Omega_X} = 0$.  However, with SUGRA turned on, and after the cosmological constant is tuned to zero, there is an explicit source of $R$ breaking and $X$ acquires a non-zero vev:
\be
\label{eq:leadingXvev}
\vev{\Omega_X} = \vev{X} =  \frac{M^2 m_{3/2}}{2 f} \left(1 + \mathcal{O}\left( \frac{M^2}{\MPl^2} \right) \right).
\ee
By adjusting the value of $M$, we can dial  $\vev{\Omega_X}$ in a controlled way.

This setup was studied in \Ref{Craig:2010yf} for $M \ll \MPl$.   Employing the naive conformal compensator $\bPhi = 1 + \theta^2 m_{3/2}$, the leading correction to the $2m_{3/2}$ goldstino mass from the vev of $X$ was found to be
\be
\label{eq:leadingdeformation}
\delta m^{(\text{vev of }X)}_\eta = -m_{3/2} \frac{M^2 m_{3/2}^2}{f^2}.
\ee
This additional correction arises because when $X$ gets a vev, $\vev{{\bOmega}^{A}|_{\theta^2}}$  in \Eq{eq:almostthere} is no longer zero and the relation in \Eq{eq:altFrelationship} no longer holds.  (See \App{app:interpolation} for further discussion.)

The new ingredient from  our improved gauge fixing is a goldstino mass term from the fermionic component of $\bPhi$.  At leading order in $\vev{X}$, one identifies an additional correction from $\vev{\Omega_i} \chi^i$ of the form
\be
\delta m^{(\text{correct}~\Phi)}_{\eta} = - \frac{\vev{\Omega_X F^X}}{3} = - \frac{1}{6} m_{3/2} \frac{M^2}{\MPl^2}.
\ee
Compared to \Eq{eq:leadingdeformation}
\be
\frac{\delta m^{(\text{correct}~\Phi)}_{\eta}}{\delta m^{(\text{vev of }X)}_{\eta}} = \frac{1}{6} \frac{f^2}{m_{3/2}^2 \MPl^2} = \frac{1}{2} \left(\frac{F_{\rm eff}^2}{F_{\rm eff}^1} \right)^2,
\ee
where we have used $f \simeq F^2_{\rm eff}$ and $m_{3/2} \simeq F_{\rm eff}^1 / \sqrt{3} \MPl$.  This ratio is small given the assumptions of our setup.

Thus, while the $\vev{\Omega_i} \chi^i$ terms in $\bPhi$ are in principle necessary to get the correct goldstino mass spectrum, we see that they can be safely ignored for small $\vev{\Omega_i}$ and when $\widetilde{F}_\Phi \simeq m_{3/2}$.  These $\vev{\Omega_i} \chi^i$ terms become important for large values of $\vev{\Omega_i}$ and $\widetilde{F}_\Phi \not= m_{3/2}$, which is the topic of the next section.

\section{Understanding Massless Modulini}
\label{sec:modulini}

The examples in \Sec{sec:goldstini} highlight the utility of the conformal compensator method together with non-linear superfield representations.   To see why the improved gauge fixing proposed in \Ref{goldstini3} is important, we want to study theories  in which $\langle \Omega_i \rangle \not= 0$.

\subsection{Almost No-Scale SUSY Breaking}

Theories of no-scale SUSY breaking \cite{Lahanas:1986uc} contain a  field $\bT$ which appears only as a linear term in the K\"ahler potential, and parameterizes the size of some extra dimension.  In these models, this no-scale field acquires a non-zero $F$-term and the cosmological constant vanishes identically.  However, the scalar component $T$ is not stabilized at tree-level.   In almost no-scale models \cite{Luty:2002hj}, $T$ is stabilized by K\"ahler dynamics,\footnote{This is in contrast to KKLT-like constructions \cite{Kachru:2003aw} where the $T$ is stabilized by non-perturbative superpotential terms.  The following discussion relies crucially on the assumption that these terms are small.} leading to a SUSY-breaking vacuum in AdS space.  By including additional ``uplifting'' SUSY-breaking sectors, one can accommodate a vanishing cosmological constant.

The structure of the minimal almost no-scale model has similarities with the minimal $F$-term goldstini construction, but some important differences.  Specifically, we assume a single no-scale field $\bT \equiv \bX^0$ comprising a sector 0 and $a= 1$ to $N$ uplifting sectors each with a single SUSY-breaking field $\bX^a$.  The action for this simple theory is given by
\begin{align}
\bOmega &= \bOmega^X +  \bOmega^T, \nonumber\\ 
\bW &= m_{3/2} + \sum_a f_a \bX^a,
\label{eq:modulino_action}
\end{align}
where
\begin{align}
\bOmega^X &= -3 + \sum_a \omega_a(\bX^{a\dagger} \bX^a),\\
\bOmega^T &= \alpha(\bT + \bT^\dagger) + \omega_0(\bT,\bT^\dagger).
\end{align}
For convenience we have shifted the no-scale field such that $\vev{T} = 0$, and performed a constant K\"ahler transformation to arrange $\Arg \vev{W} = 0$ and $\vev{\Omega} = -3$.  Here, $\alpha$ is a constant parameter, the function $\omega_0$ stabilizes $T$, and the function $\omega_a$ stabilizes the uplifting field from sector $a$.  Note that our results will hold even if $\omega_0$ explicitly breaks the shift symmetry on $T$.  Following \Sec{subsec:minimialmasscalc}, it is of course possible to include arbitrary $F$-term and $D$-term SUSY breaking in each of the $N$ uplifting sectors, in which case one obtains the same results as in this simplified example.

We see that \Eq{eq:modulino_action} is essentially a SUGRA action of the sequestered form of \Eq{eq:sequesteredSUGRA}.  However, unlike in the $F$-term and $D$-term scenario, $\vev{\Omega_T} \not= 0$ and $\vev{W_T} = 0$.  In particular, SUSY breaking in the no-scale sector depends crucially on SUGRA effects and does not even occur in the $\MPl \rightarrow \infty$ limit.  Therefore we should expect large deviations from the universal relation $m_\eta = 2 m_{3/2}$ for the goldstini masses.

\subsection{A Curious Factor of Zero}
\label{sec:curious_zero}

A deviation from the standard mass relation was derived in the appendix of \Ref{Cheung:2010mc}, where the fermionic spectrum of the almost no-scale construction in \Eq{eq:modulino_action} was calculated using the component SUGRA Lagrangian.  The tree-level spectrum of a no-scale field $\bT \equiv \bX^0$ plus $N$ Poloyni fields $\bX^a$ with $\vev{\Omega_a} = 0$ consists of:
\begin{itemize}
\item A gravitino of mass $m_{3/2}$;
\item $N-1$ fermion modes with mass $2 \widetilde{F}_\Phi \not= 2 m_{3/2}$;
\item One fermion mode with mass zero. 
\end{itemize}

The $N-1$ fermion modes are simple to understand since they correspond to uneaten goldstini which are expected to have a mass equal to $2 \widetilde{F}_\Phi$, with
\be
\widetilde{F}_\Phi \equiv \bPhi|_{\theta^2} = m_{3/2} + \frac{\vev{\Omega_T}}{3} F^T.
\ee
More surprising is the appearance of a tree-level massless mode---a modulino.
The modulino is expected to get a small mass from loops, incomplete sequestering, non-perturbative effects, or other dynamics.  Still, the fact that the modulino is massless at tree-level in the strict sequestered limit is a puzzling fact.

What makes the massless modulino particularly perplexing is that it is massless only when  two conditions are satisfied:
\begin{itemize}
\item The no-scale field is stabilized ($\partial V / \partial T = 0$);
\item The cosmological constant is tuned to zero ($V = 0$).
\end{itemize}
The first condition implies that the modulino is \emph{not} protected by a chiral symmetry, since its vanishing mass appears as a dynamical effect.  The second condition is also confusing, since $V = 0$ is not usually thought of as a symmetry enhanced point.\footnote{Indeed, one might misguidedly try to solve the cosmological constant problem by imposing a chiral symmetry on the modulino.}  When calculating fermion masses using component SUGRA methods, the masslessness arises from an unexpected cancellation, with no hint for its origin.

\subsection{Modulino as Secret Goldstino}  

Using our improved conformal compensator, we will see that the modulino can be understood as the  goldstino of an accidental \emph{global} SUSY.  That is, we can express the Lagrangian in such a way that $\bT$ exhibits an enhanced global SUSY that is spontaneously broken by an effective $F$-term vev,  $\vev{F^T} = \alpha \widetilde{F}_\Phi$.  This effect relies crucially on the $\vev{\Omega_i} \chi^i$ term in $\bPhi$, and the accidental global SUSY will only appear when $\partial V / \partial T = 0$ and $V = 0$, explaining the confusing cancellation described in \Sec{sec:curious_zero}.

The most convenient way to see this emergent SUSY is to work in unitary gauge for the gravitino.  In this gauge, $\eta_{\rm eaten}$ is projected out of the Lagrangian, and will appear as a zero eigenvalue in the fermion mass matrix.\footnote{Strictly speaking, $\eta_{\rm eaten}$ has no kinetic term in this gauge, but the zero eigenvalue will still appear as long as we start with canonically normalized kinetic terms for the fermions before going to unitary gauge.}  The eaten goldstino direction is ($i = 0$ to $N$ and $a = 1$ to $N$)\footnote{The invariant K\"ahler potential is $G = -3 \log \frac{\Omega}{-3} + \log W + \log W^*$ and $m_{3/2} = e^{\vev{G}/2}$.  See \Ref{goldstini3} for an explanation of how to identify the goldstino mode using the improved compensator method.  Normalization of these states assumes vanishing cosmological constant, i.e.\ $\vev{G_i G^i} = 3$.}
\be
\label{eq:eta_eaten}
\eta_{\rm eaten} = \frac{1}{\sqrt{3}}\vev{G_i} \chi^i = \frac{1}{\sqrt{3}}\left(\vev{\Omega_T} \chi^T + \frac{\vev{W_a} \chi^a}{m_{3/2}}  \right).
\ee
Once in unitary gauge, we will see manifestly a \emph{second} zero eigenvalue corresponding to $\eta_{\rm massless}$ which is orthogonal to $\eta_{\rm eaten}$
\be
\eta_{\rm massless} = \frac{1}{\sqrt{3}} \left( \frac{\vev{\Omega_T} \chi^T}{\gamma} - \frac{ \gamma \vev{W_a} \chi^a}{m_{3/2}}  \right), 
\ee
with
\be
\gamma \equiv \sqrt{\frac{\vev{\Omega_T \Omega^T} m_{3/2}^2}{\vev{W_a W^a}}}.
\ee
One needs to carefully account for all fermion mass mixing terms to properly separate $\eta_{\rm massless}$ from $\eta_{\rm eaten}$, and this is precisely guaranteed by the $\vev{\Omega_i} \chi^i$ term in the improved conformal compensator.  

Note that the identification of the eaten direction as \Eq{eq:eta_eaten} is only true in flat space, so the following calculations implicitly assume $V = 0$.  Furthermore, we will use a non-linear parametrization for $\bT$ in which $\bT^2 = 0$, which implicitly assumes that $T$ is stabilized at $\vev{T} = 0$.  Thus, the conditions $\partial V / \partial T = 0$ and $V = 0$ both appear as key ingredients in the derivation of the vanishing modulino mass.

\subsection{Enhanced Sequestering and Hidden SUSY}

The reason why unitary gauge is so useful for studying almost no-scale models is that the Lagrangian exhibits an enhanced ``sequestering'' of $\bs T$ which realizes an additional accidental global SUSY.  For calculating fermion masses,\footnote{If we wanted to calculate interactions, we would need to keep the full scalar field dependence in \Eq{eq:phicorrect}.} we can work with two completely equivalent versions of the conformal compensator:
\begin{align}
\bPhi^X &=1 - \sqrt{2} \theta \frac{\vev{W_a} \chi^a}{3 m_{3/2}} + \theta^2 \widetilde{F}_\Phi, \nonumber \\
\bPhi^T &=1 + \sqrt{2} \theta \frac{\vev{\Omega_T} \chi^T}{3} + \theta^2 \widetilde{F}_\Phi.
\end{align}
The difference between the fermionic components of $\bPhi^X$ and $\bPhi^T$ is $\eta_{\rm eaten}$.  Because $\vev{\Omega_a} = 0$ and $\vev{W_T} = 0$, $\bPhi^X$ only depends on the fermions in $\bX^a$ and $\bPhi^T$ only depends on the fermions in $\bT$.  The nice form of $\bPhi^X$ and $\bPhi^T$ depends crucially on the $\vev{\Omega_i} \chi^i$ terms in the conformal compensator.  

We can now write the SUGRA action as
\be
\label{eq:SUGRAdoublesequester}
\mathcal{L}_{\rm SUGRA} = \mathcal{L}^X + \mathcal{L}^T + \ldots,
\ee
where we have elided the graviton/gravitino terms and
\begin{align}
\mathcal{L}^X &= \int d^4 \theta \; \bPhi^{X\dagger} \bPhi^X \; \bOmega^X + \int d^2 \theta \; (\bPhi^{X})^3 \; \bW + \mathrm{h.c.}, \nonumber \\
\mathcal{L}^T &= \int d^4 \theta \; \bPhi^{T\dagger} \bPhi^T \; \bOmega^T.
\end{align}
As advertised, in unitary gauge $\chi^T$ and $\chi^a$ are effectively sequestered from each other.

This form also makes manifest a hidden global SUSY in $\mathcal{L}^T$.  This is most apparent using the non-linear parametrization for $\bT$ which is valid below the mass of the modulus $T$:
\be
\bT = \left(\theta + \frac{1}{\sqrt{2}}\frac{\chi^{T}}{F^{T}}\right)^2 F^T,
\ee
for which $\bT^2 = 0$ when by assumption $\vev{T} = 0$ since we are working in a basis where are fields are shifted to have zero vevs.  This leads to a dramatic simplification of $\mathcal{L}^T$, since the only terms relevant for calculating fermion masses are
\be
\bOmega^T \supset \alpha(\bT + \bT ^\dagger) + \beta \bT^\dagger \bT.
\ee
Performing a field redefinition $\bT \rightarrow \bT / \bPhi^T$, $\mathcal{L}^T$ becomes\footnote{Note that both the $\alpha$ and $\beta$ coefficients contribute to the kinetic term for $\chi^T$.}
\begin{align}
\mathcal{L}^T &= \int d^4 \theta \; \left[ \alpha \left( \bPhi^{T\dagger} \bT + \bPhi^T \bT^\dagger \right) + \beta \bT^\dagger \bT \right], \nonumber \\
& = \int d^2 \theta \; \alpha \widetilde{F}_\Phi^\dagger \bT + \text{h.c.} + \ldots.  \label{eq:Tlang}
\end{align}
Thus, the non-linear form of $\bT$ behaves like a chiral multiplet that breaks a hidden global SUSY, since the $\bT$ equation of motion yields $\vev{F^T} = \alpha \widetilde{F}_\Phi$.  

Observe that at arbitrary points in field space there will be mass terms for $\chi^T$.  It is only at the minimum of the potential, $\vev{T} = 0$, that this mass term vanishes, and the non-linear parametrization is helpful for extracting physics in the vicinity of $\partial V / \partial T = 0$.  

\subsection{The Final Result}

We now have the ingredients to calculate the spectrum of fermions from \Eq{eq:SUGRAdoublesequester}.  From $\mathcal{L}^T$, the field $\chi^T$ acts like  a massless ``goldstino'' for the hidden global SUSY, implying that the spectrum in unitary gauge contains:
\begin{itemize}
\item One zero eigenvalue in the $\chi^T$ direction.
\end{itemize}
Morover, we recognize $\mathcal{L}^X$ as being identical to the Lagrangian in the minimal goldstino scenario, albeit with $\widetilde{F}_\Phi \not= m_{3/2}$.  We have already solved this system in \Sec{subsec:minimialmasscalc} in non-unitary gauge, where we found that all of the fermions have mass $2 \widetilde{F}_\Phi$.  Now in unitary gauge, this part of the Lagrangian yields:
\begin{itemize}
\item One zero eigenvalue in the $W_a \chi^a$ direction,
\item $N-1$ eigenvalues equal to $2 \widetilde{F}_\Phi$.
\end{itemize}
Note that the $W_a \chi^a$ direction corresponds to the would-be goldstino direction in the absence of $\mathcal{L}^T$.  Because both $W_a \chi^a$ and $\chi^T$ are massless directions, we can rotate this subsystem to identify the eaten goldstino $\eta_{\rm eaten}$ and the massless modulino $\eta_{\rm massless}$.  Since we are in unitary gauge already, $\eta_{\rm eaten}$ does not correspond to a physical degree of freedom, but  $\eta_{\rm massless}$ persists as the advertised accidentally massless modulino.

It is now clear the origin of the massless modulino.  The almost no-scale model in \Eq{eq:modulino_action} has two levels of sequestering: a sequestering among the $\bX^a$ which is broken only by $\widetilde{F}_\Phi$, and an additional sequestering between the $\bX^a$ and $\bT$ which occurs when $\partial V / \partial T = 0$ and $V = 0$.  The fermions $W_a \chi^a$ and $\chi^T$ are effectively decoupled in unitary gauge, and the massless modulino appears because $\chi^T$ behaves as if it were a ``goldstino'' for its own hidden global SUSY.

Obviously, this massless modulino will become massive if $\vev{W_T} \not= 0$ or $\vev{\Omega_a} \not=0$, or if there were direct couplings between $\bX^a$ and $\bT$ in the SUGRA action.  Indeed, for a phenomenologically viable almost no-scale model, one would likely need to lift this massless mode.

\section{Conclusions}
\label{sec:conclusions}

Much of  SUGRA literature has focused on the scalar spectrum of SUGRA theories.  However, surprises can appear in the fermionic spectrum which can substantially impact phenomenology.  In this paper, we have clarified two interesting features of goldstini spectra in theories of multiple SUSY breaking:  the  factor of two in the $m_\eta = 2 m_{3/2}$ goldstini mass relation, and the curious vanishing mass of the modulino in almost no-scale models.  

A clear understanding of these results is difficult in the standard component formulation of SUGRA.  We have seen that the fermionic spectrum is much easier to understand via calculations performed directly in superspace, and the improved conformal compensator method introduced in our companion paper \cite{goldstini3} has made these calculations possible without worrying about complications from graviton/gravitino mixing.  At the very minimum, the formalism of \Ref{goldstini3} shows under what circumstances the naive $\bPhi \simeq 1 + \theta^2 F_\Phi$ parametrization is valid and when one must account for $\vev{\Omega_i} \chi^i$ terms in the conformal compensator.

Beyond the spectrum of fermions, one is also interested in the interactions of fermions with other fields, and we expect that the improved compensator formalism will help clarify issues concerning these as well.  As one example which we will pursue in future work, recall the goldstino equivalence theorem, which states that at high energies, the couplings of a matter multiplet to the helicity-1/2 components of the gravitino can be described by the coupling to the goldstino.  This is readily apparent in models where SUSY breaking is communicated to the standard model fields $\bQ$ via a SUSY breaking multiplet $\bX$, since operators such as $\bX^\dagger \bX \bQ^\dagger \bQ / \Lambda^2$ not only generate soft masses for the sfermions, but also contain the desired goldstino-sfermion-fermion coupling.  However, if SUSY breaking is communicated to standard model fields by the conformal compensator $\bPhi$ as in anomaly mediation \cite{anomalymediation}, there is a mismatch between the soft mass term and the goldstino-sfermion-fermion coupling, since the fermionic component of $\bPhi$ only contains the $\vev{\Omega_i} \chi^i$ part of the goldstino.  We expect the improved compensator method will help clarify this apparent violation of the goldstino equivalence theorem.

\begin{acknowledgments}
We thank Markus Luty, Yasunori Nomura, Raman Sundrum, and Jay Wacker for helpful conversations.  C.C is supported in part by the Director, Office of Science, Office of High Energy and Nuclear Physics, of the US Department of Energy under Contract DE-AC02-05CH11231 and by the National Science Foundation on grant PHY-0457315.   F.D. and J.T. are supported by the U.S. Department of Energy under cooperative research agreement Contract Number DE-FG02-05ER41360.  

\end{acknowledgments}

\appendix

\section{Details of Goldstini Calculation}
\label{app:details}

In \Sec{subsec:minimialmasscalc}, we derived the universal tree-level goldstino mass of $2m_{3/2}$ by using a convenient parametrization of the chiral multiplets
\be
\label{eq:nonlinearX}
\bX^{Ai} = X^{Ai} + \left(\theta + \frac{1}{\sqrt{2}}  \frac{\eta^A}{F^A_{\rm eff}}\right)^2 F^{Ai},
\ee
where we have reinstated the scalar component $X^{Ai}$, and are still only considering the goldstino direction $\eta^A$.  There is a similar expression for the vector multiplets if one fixes to the analog of Wess-Zumino gauge.

While it is possible to derive the universal tree-level mass using the standard linear parameterization
\be
\label{eq:linearX}
\bX^{Ai} = X^{Ai} + \sqrt{2}\theta \eta^A \frac{F^{Ai}}{F^A_{\rm eff}} + \theta^2 F^{Ai},
\ee
the derivation becomes far more cumbersome.  For example, if one were to use the linear parametrization in the model from \Eq{eq:BXLagrangian}, the universal $2m_{3/2}$ goldstini mass would comes from the $(\bX^\dagger \bX)^2/M^2$ term after inserting the $X$ vev, but taking $\vev{X} \rightarrow 0$ to achieve $\vev{\Omega_X} = 0$.

It is clear that \Eq{eq:nonlinearX} and \Eq{eq:linearX} are related by a simple field redefinition on the scalar component
\be
X^{Ai} \rightarrow X^{Ai} + \eta^A \eta^A  \frac{F^{Ai}}{2 (F^A_{\rm eff})^2}.  
\ee
Thus, either the linear or non-linear parameterization of the chiral multiplet is fine for calculational purposes.  The reason we prefer \Eq{eq:nonlinearX} is not only that it simplifies the calculation, but it has the physical interpretation of performing a broken SUSY transformation on the vacuum,
\be
\bX^{Ai} = \exp \left[ \frac{Q \eta^A}{\sqrt{2} F^A_{\rm eff}}  \right] \left( X^{Ai} + \theta^2 F^{Ai}  \right),
\ee
where $Q_\alpha \equiv \partial/\partial \theta^\alpha$ is a generator of SUSY transformations.  This is analogous to the convenient parametrization of a Higgs field as $e^{i \pi / \sqrt{2} f} (f + h/\sqrt{2})$.

An important ingredient to deriving $2m_{3/2}$ was \Eq{eq:altFrelationship}, repeated for convenience
\be
{\bOmega}^{A}|_{\theta^4} + 3 {\bW}^{A}|_{\theta^2} = - 2 (F_{\rm eff}^{A})^2.
\tag{\ref{eq:altFrelationship}}
\ee
It is easy to understand why this is true by expanding out the multiplets
\begin{align}
{\bOmega}^{A}|_{\theta^4} &= \frac{D^{Aa} d^{Aa}}{2} + F^{Ai} F^{*A\bar{j}} g_{i\bar{j}}, \\
 {\bW}^{A}|_{\theta^2} &= F^{Ai} f^{Ai} ,
\end{align}
where $d^{Aa}(X^{\dagger A\bar{j}}, X^{Ai})$ and $f^{Ai}(X^{Ai})$ are functions of the scalar fields.  The $F$- and $D$-term equations of motion set
\begin{align}
f^{Ai} &= - F^{A\bar{j}} g_{i\bar{j}}, \\
d^{Aa} & = - D^{Ab} f^A_{ab}.
\end{align}
So using the relationship $1-3 = -2$, we indeed recover \Eq{eq:altFrelationship} using the definition of $F_{\rm eff}^{A}$ in \Eq{eq:goldstinodecayconstant}.

\section{Leading Deformation}
\label{app:interpolation}

In \Sec{sec:goldstinideformations}, we studied the leading deformation from the universal $2m_{3/2}$ result when $\vev{\Omega_i} \not= 0$.  We argued that the $\vev{\Omega_i} \chi^i$ term in $\bPhi$ gave a subdominant contribution to the goldstino mass compared to the leading effect from \Eq{eq:leadingdeformation}.  In this appendix, we want to understand the leading deformation in more detail.

The easiest way to proceed is to go back to \Eq{eq:almostthere}, where the $\vev{\Omega_i} \chi^i$ terms were ignored and $F_\Phi = m_{3/2}$.   In that limit, the goldstino mass for sector $A$ is
\be
\label{eq:genericmass}
m_\eta = - m_{3/2} \left( \frac{ m_{3/2} {\bOmega}^{A}|_{\bar{\theta}^2} + {\bs \Omega}^{A}|_{\theta^4} + 3 {\bs W}^{A}|_{\theta^2}}{(F_{\rm eff}^{A})^2} \right).
\ee
For generic parameter values for the model in \Eq{eq:BXLagrangian}, $\vev{X}$ can be substantial, so $\vev{{\bOmega}^{A}|_{\theta^2}}$ is no longer zero and \Eq{eq:altFrelationship} no longer holds.  However, \Eq{eq:genericmass} is still true as long as we are in the limit where $F^1_{\rm eff} \gg F^2_{\rm eff}$.

\begin{figure}[t]
\begin{center}
\includegraphics[scale=1.3]{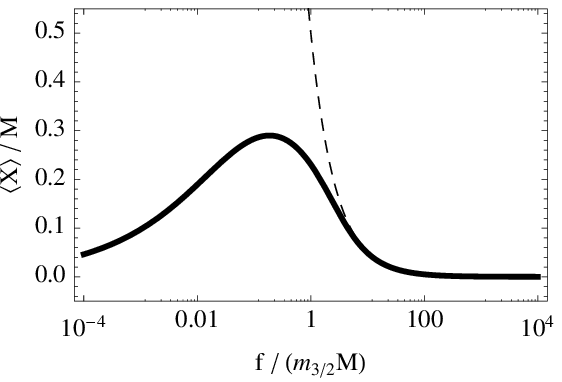}
\end{center}
\caption{The value of $x \equiv \vev{X}/M$ as a function of $\epsilon \equiv f/m_{3/2}M$ for the Lagrangian in \Eq{eq:BXLagrangian}.  The dashed line corresponds to the approximate solution in \Eq{eq:leadingXvev}.}
\label{fig:Xvev}
\end{figure}

\begin{figure}[t]
\begin{center}
\includegraphics[scale=1.3]{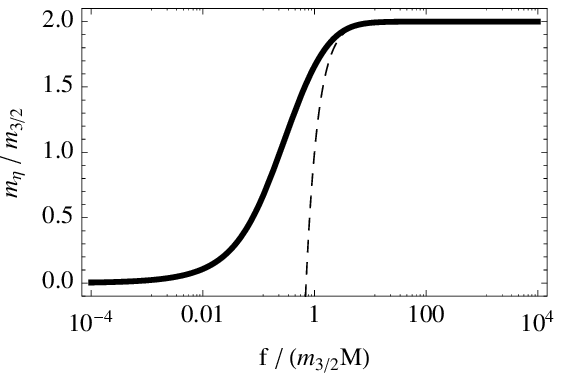}
\end{center}
\caption{Mass of the deformed goldstino from \Eq{eq:BXLagrangian} as a function of $\epsilon = f/m_{3/2}M$.  When $\epsilon$ is large, this corresponds to large SUSY breaking and a well-stabilized sgoldstino, and one recovers the universal goldstino mass of $2m_{3/2}$.  When $\epsilon$ is small, this corresponds to small SUSY breaking and a poorly stabilized sgoldstino, and the goldstino becomes less massive.  The dashed line indicates the approximation in \Eq{eq:leadingdeformation}.}
\label{fig:interpolation}
\end{figure}

From the Lagrangian in \Eq{eq:BXLagrangian}, and only considering sector 2, we have
\be
{\bOmega}^{A=2}|_{\theta^2}  = \Omega_X F^X,  \quad {\bOmega}^{A=2}|_{\theta^4}  = \Omega_{X^\dagger X} |F^X|^2, 
\ee
\be
{\bW}^{A=2}|_{\theta^2}  = f F^X , \qquad F^{A=2}_{\rm eff} = F^X \sqrt{\Omega_{X^\dagger X}}.
\ee
The $F^X$ equation of motion yields
\be
F^X = - \frac{f + \Omega_{X^\dagger} m_{3/2}}{\Omega_{X^\dagger X}},
\ee
so we can write the fermion mass in \Eq{eq:genericmass} as
\be
m_{\eta} = 2 m_{3/2} \left(1 + \frac{m_{3/2} \Omega_{X^\dagger}}{ F^{*X} \Omega_{X^\dagger X} } \right).
\ee
Putting in the explicit form for $\Omega$ and $F^{X}$
\be
m_{\eta} = 2 m_{3/2} \frac{\epsilon}{\epsilon + x - 2 x^3},
\ee
where we have defined
\be
\qquad \epsilon \equiv \frac{f}{m_{3/2} M}, \qquad x \equiv \frac{\vev{X}}{M}.
\ee
The numerical solution for $x$ as a function of $\epsilon$ appears in \Fig{fig:Xvev}.

Large $\epsilon$ corresponds to small $M$, which means that $X$ is well-stabilized near zero.  Not surprisingly, we recover the universal goldstino mass of $2m_{3/2}$ in that limit.  As $\epsilon$ decreases, then the goldstino get correspondingly lighter, with the precise mass depending on the vev of $X$.   The full interpolation from a massless fermion that provides no SUSY breaking ($f = 0$) to a goldstino with the universal mass ($\epsilon \rightarrow \infty$) is shown in \Fig{fig:interpolation}.

\end{document}